\RequirePackage{fix-cm}
\documentclass[epjc3]{svjour3}  
\RequirePackage{color}
\smartqed  
\RequirePackage{graphicx}
\usepackage{lineno}
\usepackage{hyperref}  
\usepackage{xcolor}
 
\journalname{Eur. Phys. J. C}

\usepackage{soul}

\begin{document}

\title{Background reduction at the KATRIN experiment by the shifted analysing plane configuration}


\author{Alexey Lokhov \thanksref{e1,addr1,addr2}          
        \and 
        Benedikt Bieringer \thanksref{addr1}
        \and 
        Guido Drexlin \thanksref{addr31, addr3}
        \and 
        Stephan Dyba \thanksref{addr1}
        \and 
        Kevin Gauda \thanksref{addr1}
        \and 
        Florian Fr\"ankle \thanksref{addr3}
        \and 
        Ferenc Gl\"uck \thanksref{addr3}
        \and 
        Volker Hannen \thanksref{addr1}
        \and
        Dominic Hinz \thanksref{addr31,addr3}
        \and 
        Susanne Mertens \thanksref{addr4, addr5}
        \and 
        Caroline Rodenbeck \thanksref{addr1}
        \and
        Anna Schaller \thanksref{addr4, addr5}
        \and 
        Christian Weinheimer\thanksref{addr1} 
}

\thankstext{e1}{E-mail: alexey.lokhov@wwu.de}


\institute{
Institut f\"ur Kernphysik, Westf\"alische Wilhelms-Universit\"at M\"unster, Wilhelm-Klemm-Str. 9, 48149 M\"unster, Germany \label{addr1}
\and
Institute for Nuclear Research of Russian Academy of Sciences, 60th October Anniversary Prospect 7a, 117312 Moscow, Russia \label{addr2}
           \and
           Institute of Experimental Particle Physics (ETP), Karlsruhe Institute of Technology (KIT), Hermann-von-Helmholtz-Platz 1, 76344 Eggenstein-Leopoldshafen, Germany \label{addr31}
           \and
            Institute for Astroparticle Physics (IAP), Karlsruhe Institute of Technology (KIT), Hermann-von-Helmholtz-Platz 1, 76344 Eggenstein-Leopoldshafen, Germany \label{addr3}
            \and
            Technische Universit\"at M\"unchen, James-Franck-Str. 1, 85748 Garching, Germany \label{addr4}
           \and
            Max-Planck-Institut f\"ur Physik, F\"ohringer Ring 6, 80805 M\"unchen, Germany \label{addr5}
}

\date{Received: date / Accepted: date}

\maketitle

\begin{abstract}
The KATRIN experiment aims at measuring the electron neutrino mass with a sensitivity of 0.2\,eV$/c^2$ after five years of data taking. Recently a new upper limit for the neutrino mass of 0.8\,eV$/c^2$ (90\% CL) was obtained. To reach the design sensitivity, a reduction of the background rate by one order of magnitude is required. The shifted analysing plane (SAP) configuration exploits a specific shaping of the electric and magnetic fields in the KATRIN main spectrometer to reduce the spectrometer background by a factor of two. We discuss the general idea of the SAP configuration and describe the main features of this novel measurement mode.
\end{abstract}

\section{Introduction}
\label{intro}
\label{sec:1}

The three neutrinos are the only particles in the standard model of particle physics with unknown absolute masses. The difficulty in measuring the neutrino masses comes from the fact that they are more than five orders of magnitude smaller than the mass of an electron and that neutrinos are neutral particles participating in weak interactions only. 
Since neutrino oscillation experiments prove that neutrinos have a mass, but only allow determining differences between the squared masses of the different neutrino mass eigenstates, other methods need to be applied. To assess the sum of the three neutrino masses, cosmological observables, mainly from the cosmic microwave background and baryon acoustic oscillation measurements, are investigated \cite{ref:Planck}. Searches for neutrinoless double beta decay provide a way to define whether the neutrino is a Dirac or Majorana fermion and are also able to determine an effective mass value $\langle m_{\beta\beta}\rangle = |\sum U_{ei}^2 m_i|$ if neutrinos are Majorana particles~\cite{ref:CUORE,ref:GERDA,ref:EXO,ref:Kamland}. Here $U_{ei}$ are the mixing matrix elements and $m_i$ are the masses of the corresponding eigenstates. Both aforementioned methods are, however, model-dependent.
Direct neutrino mass measurements exploit the kinematics of weak processes ($\beta$-decay, electron capture) enabling a model-independent determination of the effective electron neutrino mass squared $ m_{\nu}^2 = \sum |U_{ei}|^2 m_i^2$~\cite{ref:Formaggio2021,ref:Drexlin2013}.

Such kinematics-based measurements study the shape of the energy spectrum of $\beta$-decay or electron capture processes and require high statistics and high energy resolution combined with very low experimental background rates.
These requirements are fulfilled in the KATRIN experiment, which has recently obtained a new upper limit on the neutrino mass of 0.8\,eV (90\% CL) after its first two neutrino mass measurement campaigns \cite{ref:KNM1PRL,ref:KNM1PRD,ref:KNM2}. 

The ultimate sensitivity of KATRIN of 0.2\,eV (90\% CL) within 3 years of total measurement time requires a background rate of ${\cal O}(10)$\,mcps \cite{ref:TDR} while -- after multiple rounds of improvements -- the current background rate of 220\,mcps with 117 out of 148 pixels \cite{ref:KNM2} still exceeds this requirement by more than an order of magnitude. Applying the simple scaling laws given in \cite{ref:TDR} (footnote 34) and in  \cite{ref:Ott08} (equation 48), the statistical sensitivity to the neutrino mass is reduced due to the higher background by about a factor of 1.4. Therefore, additional background reduction measures are required. An effective approach using the so-called shifted analysing plane (SAP) configuration is discussed in this paper and allows to reduce the background by about a factor two w.r.t. the nominal operating mode.

This paper is structured as follows. Section~\ref{sec:2} summarizes briefly the KATRIN experiment, focusing on the working principle of the main spectrometer. In section~\ref{sec:3} KATRIN's main background sources are presented. In section~\ref{sec:4} the new SAP configuration is explained and characterised via the electromagnetic field simulations and first measurements of the background rate in this mode. Section~\ref{sec:new} discusses briefly the implications of the new measurement mode to the neutrino mass measurements. Our conclusions are given in section~\ref{sec:5}.

\section{KATRIN's working principle}
\label{sec:2}
The KATRIN experiment uses a high-luminosity gaseous molecular tritium source ($10^{11}$~Bq) combined with a high-resolution electron spectrometer providing an energy resolution of ${\cal O}(1)$\,eV. The beta decay electrons are guided by strong magnetic fields towards the spectrometer while the tritium flow is reduced by 14 orders of magnitude using differential and cryogenic pumping techniques~\cite{ref:DPS}. The spectrometer of MAC-E-Filter (Magnetic Adiabatic Collimation with Electrostatic Filter) type acts as a high pass filter for the electrons. By scanning the retardation voltage of the spectrometer, the filter energy is varied around the tritium beta endpoint of 18.6\,keV. The electrons that are transmitted by the spectrometer are counted by a 148-pixel
silicon PIN-diode focal plane detector \cite{ref:FPD}.
\begin{figure*}
 \centering
 \includegraphics[width=0.9\textwidth]{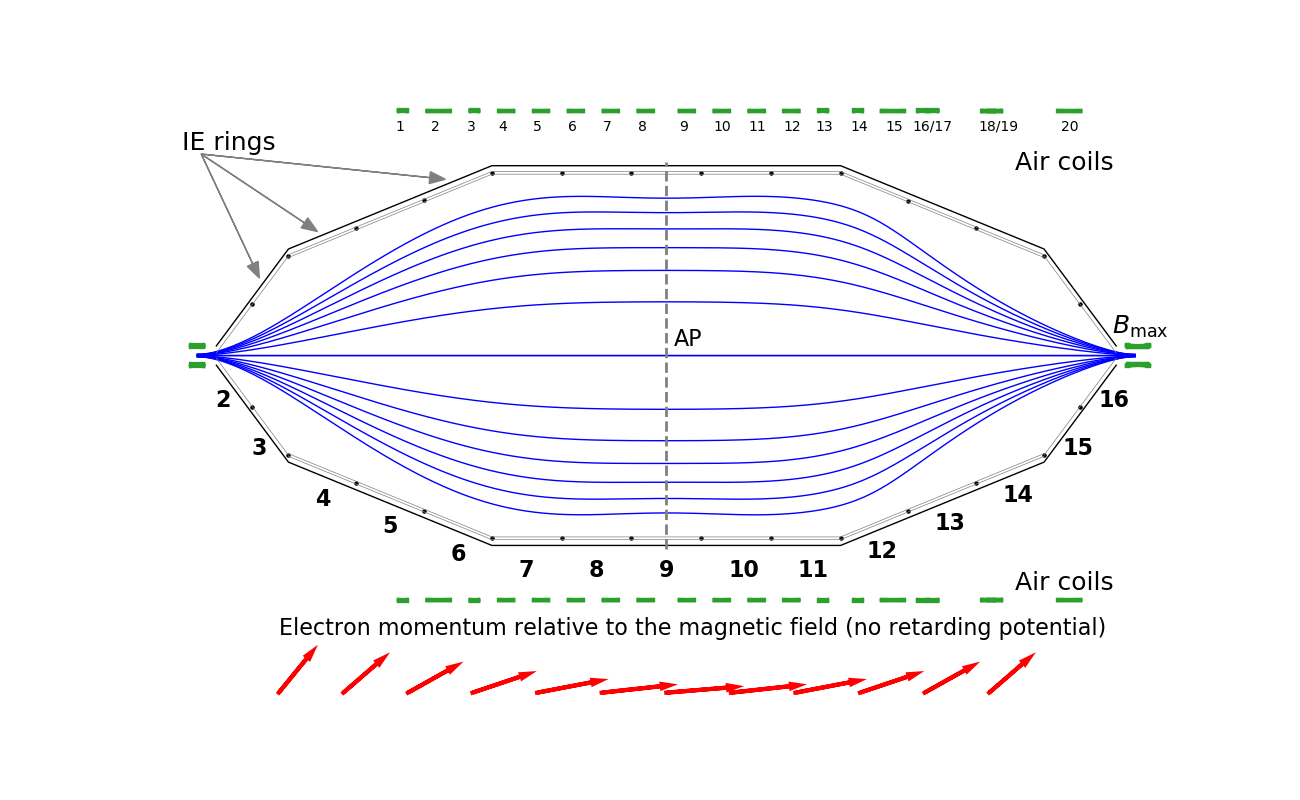}
 \caption{The KATRIN main spectrometer with its system of air coils and inner electrodes and illustration of the MAC-E-Filter principle. Electrons enter the spectrometer from the left and follow the magnetic field lines (solid, blue) towards the analysing plane (AP, dashed, gray). Due to orbital magnetic moment conservation the momenta of electrons are aligned with the magnetic field lines in the analysing plane region (red solid arrows in the bottom). The magnetic field is shaped by the superconducting solenoids at the entrance and the exit of the spectrometer and the system of air coils (numbers 1 to 20 are given in the top part of the scheme). The maximum magnetic field $B_\mathrm{max}$ is produced by the pinch magnet at the exit of the spectrometer. The electric potential is applied to the vessel and the inner electrodes (IE). Typically the IE are at about 200 V more negative potential than the spectrometer vessel. The IE rings (numbers 2-16 shown at the bottom part) are used to shape the electric potential. } 
 \label{fig:0}
\end{figure*}

In order to understand the new SAP background reduction method, we have to discuss the working principle of the KATRIN spectrometer in more detail, see Fig.~\ref{fig:0}. On the way from the spectrometer entrance (left in Fig.~\ref{fig:0}) to the analysing plane (dashed gray line, AP), where the electric retarding potential $qU$ for a particle of a charge $q$ and a retarding voltage $U$ reaches its maximum, the magnetic field, created by two superconducting solenoids at the entrance (left) and the exit (right) of the spectrometer, is reduced by four orders of magnitude (solid blue lines in Fig.~\ref{fig:0} show several magnetic field lines). The electrons are guided adiabatically by the magnetic field leading to a constant orbital magnetic moment $\mu$, which is (in the non-relativistic limit) given by the field strength $B$ and the energy due to transverse motion $E_\perp$
\begin{equation}
  \label{eq:conservation_orbital_momentum}
  \mu = \frac{E_\perp}{B} = \mathrm{const}.
\end{equation}
Therefore, the momenta of the electrons become aligned ($E_\parallel \gg E_\perp$) with the magnetic field lines in the area of the weak magnetic field (lower part of Fig.~\ref{fig:0}, red solid arrows), while the total kinetic energy $E=E_\perp + E_\parallel$ is conserved. 
Because of the conservation of orbital magnetic moment, see Eq. (\ref{eq:conservation_orbital_momentum}), the maximum electron energy due to transverse motion $E_\perp$ in the analysing plane and therefore the width of the transmission function (energy or filter width) of the spectrometer amounts to:
\begin{equation}
 \label{eq:energyWidth}
  \Delta E = \frac{B_{\mathrm{min}}}{B_{\mathrm{max}}} \cdot E= \frac{B_{\mathrm{ana}}}{B_{\mathrm{max}}} \cdot E \; ,
\end{equation}
where $B_{\rm ana}$ labels the magnetic field in the analysing plane and $B_{\rm max}$ the maximum magnetic field encountered in the beamline, which is produced by the pinch magnet at the exit of the spectrometer, see Fig.~\ref{fig:0}.

The electric retarding potential $qU$ superimposed on the magnetic field reduces the kinetic energy of the electrons and reflects the ones with an insufficient starting kinetic energy. The maximum of the retarding potential and the minimum of the absolute magnetic field define the above-mentioned analysing plane -- a virtual surface where the electrons with different starting pitch angles $\vartheta$ to the magnetic field lines have the smallest $E_\parallel$, the energy due to the motion in the direction of the magnetic field (solid blue magnetic field lines in Fig.\ref{fig:0}).

The electric potential profile within the KATRIN main spectrometer is defined by the voltages applied to the spectrometer vessel and the inner electrodes (IE) that consist of several rings of wire modules lining the inside of the vessel and allow for fine-tuning of the electric retarding potential inside the spectrometer \cite{ref:IE,ref:TDR2}. The magnetic flux tube is shaped by the superconducting magnets of the KATRIN beam line and a system of air coils around the main spectrometer \cite{ref:FG2013,ref:AC}. The system of 20 air coils along the beam axis can provide an axially symmetric magnetic field in the range of 0-2\,mT and allows for a fine-tuning of the magnetic field at various positions along the axis.
In the nominal configuration the analysing plane is located in the middle of the spectrometer (dashed gray line, Fig.\ref{fig:0}). Using different currents of the air coils and different potentials applied to the IE, the analysing plane can be moved to a different position along the beam axis.

\section{KATRIN background sources}
\label{sec:3}
The main sources of KATRIN background were extensively studied over the last years. All except one background source are successfully mitigated using passive or active countermeasures.

Background events induced in the vessel walls by cosmic muons are efficiently suppressed by electrostatic and magnetic shielding. The electrostatic shielding is provided by a potential difference of -200\,V between the inner electrodes and the spectrometer vessel. Electrons produced near the spectrometer walls are also reflected by the Lorentz force back to the walls \cite{ref:muon}.
With active electrostatic and magnetic shielding the external gamma radioactivity has been proven to make only a minor contribution to the total background rate of KATRIN \cite{ref:gamma}, at less than $5$\,mcps.

A contribution of the Penning trap between the main spectrometer and the pre-spectrometer to the background rate was studied in detail and a dedicated electron catcher system is used to remove trapped electrons from the inter-spectrometer region \cite{ref:penning}. To avoid time-dependent background rate of the Penning trap \cite{ref:KNM1PRD,ref:KNM2} the pre-spectrometer was grounded in the later measurement campaigns.
The focal-plane detector of KATRIN contributes $\sim35$\,mcps to the background rate due to radioactivity and cosmic muons. This background is reduced to $\sim12$\,mcps by applying stricter energy cuts and by using a muon veto \cite{ref:FPD}.

A part of the background events is caused by short-lived $^{220}$Rn and $^{219}$Rn atoms emanating from the material of non-evaporable getter (NEG) pumps mounted inside the pump-ports of the spectrometer and from the walls of the main spectrometer. The neutral radon atoms propagate into the spectrometer volume and decay producing primary electrons with energies in the range of $10^1-10^5$\,eV \cite{ref:wandkowski}.
The high energy electrons are magnetically trapped between the two superconducting solenoids at the entrance and the exit of the spectrometer. They scatter on the residual gas in the volume, thereby producing secondary electrons that may eventually reach the detector. 
The radon-induced background component is substantially suppressed by introducing LN$_2$-cooled copper baffles in front of the NEG pumps which prevent radon atoms from reaching the spectrometer volume \cite{ref:goerhardt}.
The baffle system has been shown to have about $95$\,\% efficiency \cite{ref:Harms}, not quite eliminating this background. The radon-induced background counts do not follow a Poisson distribution due to the time correlation between the secondary events in a cluster of counts produced in a short time by the same primary electron. This non-poissonian behaviour leads additionally to an effective increase of statistical fluctuations and reduces the sensitivity of the neutrino mass measurement \cite{ref:KNM1PRL,ref:KNM1PRD,ref:KNM2}.

The remaining and thus now dominating component of the main spectrometer background is thought to be related to $\alpha$-decays in the spectrometer walls, causing neutral particles to propagate into the spectrometer volume \cite{ref:Rydberg}. Under this hypothesis the highly excited (Rydberg) states of atoms (mostly hydrogen) are sputtered from the walls due to $\alpha$-decays of $^{210}$Po, the daughter isotope of the long-lived isotope $^{210}$Pb ($t_{1/2} = 22.2$\,yr) that had been implanted in the spectrometer vessel walls by $^{222}$Rn decays during the installation of the inner electrode inside the spectrometer under ambient air circulation. 
The $\alpha$-decay of $^{210}$Po produces $^{206}$Pb recoil ions with high momentum, which can sputter off atoms from the walls creating atoms in excited and ionised states and, thus, also Rydberg states.  
Being neutral, these atoms can propagate into the volume of the spectrometer. 
There is a certain probability for the Rydberg states to be ionised by thermal radiation from the vessel walls which are kept at a nominal temperature of $T \approx 10\,^\circ$C.
This mechanism produces electrons with energies of the order of $k_\mathrm{B}T \approx 25$\,meV, nearly homogeneously distributed within the main spectrometer volume.
These ``Rydberg electrons'' are accelerated by the electric field of the main spectrometer either towards the source or towards the detector, depending whether they are created upstream or downstream of the analysing plane, see Fig.~\ref{fig:1}.
The Rydberg electrons produced near the analysing plane will be accelerated to a kinetic energy of $qU_\mathrm{AP}$ with $U_\mathrm{AP}$ being the retarding potential in the analysing plane. Therefore, those Rydberg electrons are energetically indistinguishable from the detected signal electrons, which have energies of $qU_\mathrm{AP}$ and slightly above in the scans of the $\beta$-spectrum.

Assuming a homogeneous distribution of Rydberg atoms within the volume of the main spectrometer, the number of background electrons produced via the ionisation of the Rydberg atoms and counted at the detector is proportional to the volume of the ``downstream flux tube'', the volume of the flux tube between analyzing plane and detector, that is mapped to the detector (see Fig.~\ref{fig:1}).

The next section describes a novel configuration of the electromagnetic field that leads to a substantial reduction of the downstream flux tube volume and therefore a reduction of the background electron rate of the KATRIN main spectrometer.


\section{Shifted analysing plane configuration}
\label{sec:4}

The total magnetic flux of 134\,T$\cdot$cm$^2$ transported from the tritium source to the spectrometer and the detector of KATRIN is a conserved quantity along the beamline and the profile of the magnetic field (the component in the direction of the spectrometer central axis $z$) defines the maximal radius of the flux tube at any position in the spectrometer.
The basic idea of the shifted analysing plane configuration is to reduce the volume-dependent background rate from Rydberg electrons by reducing the effective volume of the downstream flux tube in the main spectrometer. 

\begin{figure*}
 \centering
 \includegraphics[width=0.9\textwidth]{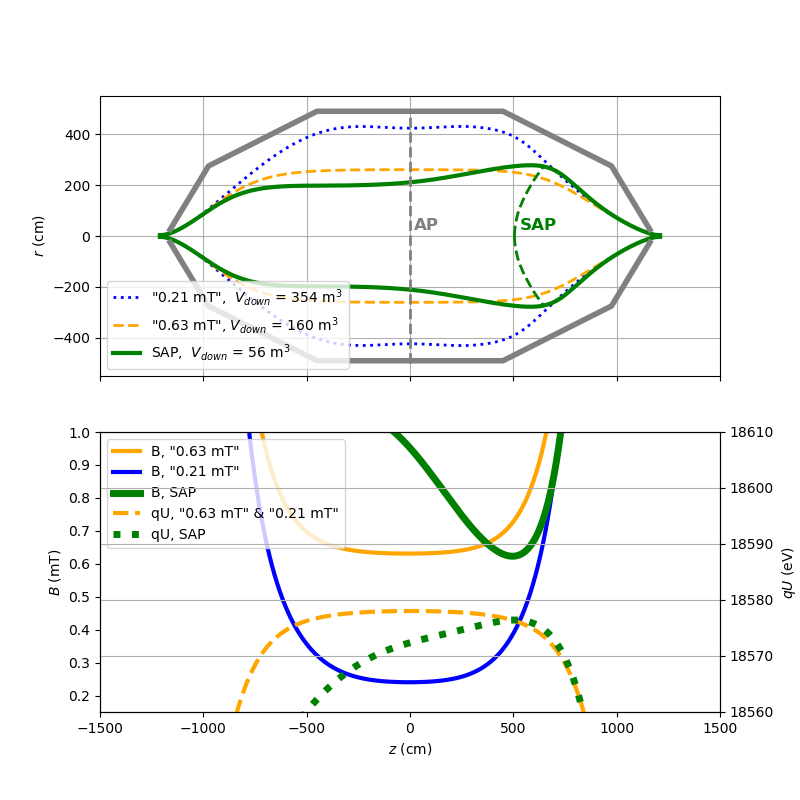}
 \caption{Upper panel: KATRIN main spectrometer with magnetic flux tube for different magnetic field settings, $V_\mathrm{down}$ gives the ``downstream flux tube'' volume. The actual position of the analysing plane is shown by the dashed lines: gray for the nominal analysing plane (AP) and green for the SAP. Lower panel: Electric retarding potential and magnetic field (both on the symmetry axis for $r=0$) for the same settings: KATRIN design report setting  (blue, \cite{ref:TDR}), symmetric but increased magnetic field setting (orange, \cite{ref:TDR2}), SAP setting (green, this work).} 
 \label{fig:1}
\end{figure*}

During the first commissioning run of KATRIN \cite{ref:FT} the magnetic field in the analysing plane was increased from 0.21\,mT (Fig.\,\ref{fig:1}, blue) to 0.63\,mT (Fig.\,\ref{fig:1}, orange) with the help of the air coil system surrounding the spectrometer vessel \cite{ref:Erhard18} in order to reduce the volume of the downstream flux tube from 341\,m$^3$  to 160\,m$^3$ and, thus, the  background rate from about 0.6\,cps to a level of about 0.3\,cps. This setting was used for the first neutrino mass measurements of KATRIN \cite{ref:KNM1PRL,ref:KNM1PRD,ref:KNM2}. 

The set point of the magnetic field (0.63\,mT) in the analysing plane was chosen to balance the statistical uncertainty gain coming from the reduced spectrometer background and the uncertainty increase due to a worse filter width of the spectrometer at higher magnetic fields in the analysing plane, see Eq.~\ref{eq:energyWidth}. The filter width is reduced to $\Delta E = 2.8$\,eV compared to the design value of 0.93\,eV in the nominal setting \cite{ref:TDR}. An additional constraint comes from the systematic uncertainties of the knowledge of the magnetic fields, derived from a difference between the simulated magnetic field and the one measured by a system of magnetometers \cite{ref:Let18,ref:TDR2}.  

In the nominal configuration of KATRIN the analysing plane is placed in the center of the main spectrometer. This setting provides several advantages \cite{ref:TDR,ref:TDR2}. It allows for the largest possible diameter of the magnetic flux tube and therefore for a small relative filter width $\Delta E / E$ and for a small spatial variation of the fields. 
The nominal configuration provides very shallow minima of the magnetic field and retarding potential stretching over the central part of the spectrometer.  
Even more importantly, in such a symmetric configuration, the minimum of the magnetic field and the maximum of the absolute retarding potential coincide with the minimum of $E_\parallel$ - the electron's kinetic energy due to motion in the direction of the magnetic field - for all radii $r$ and regardless of the start pitch angle of the electron in the source $\vartheta_\mathrm{start}$\,\cite{ref:FG2013}. If the minima are not the same for electrons with different pitch angles $\vartheta_\mathrm{start}$, the transmission function will get broadened  \cite{ref:TDR2} up to the unwanted effect of ``too early retardation'' \cite{ref:Valerius}. That effect occurs if the minima of $E_\parallel$ and $B_\mathrm{min}$ do not coincide so that the electrons reach the minimum of $E_\parallel$ with a higher residual transverse kinetic energy while the collimation of the electrons' momenta occurs at a different position with the minimal magnetic field.

With the increase of the magnetic field in the analysing plane from 0.21\,mT \cite{ref:FG2013} to 0.63\,mT, the maximum radius of the transported flux tube is reduced from 4.5\,m to 2.6\,m and therefore the position of the analysing plane at the center of the spectrometer is no longer strongly constrained by the spectrometer geometry (see Fig. \ref{fig:1}).
Therefore, we can even go a step further. The segmented inner electrode system of the KATRIN main spectrometer and the large number of air coil magnets around the vessel allow us to shape the fields in a different way: both the position of the maximal absolute retarding potential and that of the minimal magnetic field can be shifted along the $z$-axis of the spectrometer. In the SAP configuration the analysing plane is shifted towards the detector section. 
Despite the change of the layout of the electromagnetic field, the transmission properties of the KATRIN main spectrometer, high acceptance ($0-51^\circ$) and narrow filter width ($\Delta E<2.8$\,eV) as well as the axial symmetry are preserved in the SAP configuration. The analysing plane is not located at a constant $z$-value but is a curved surface, see Fig.~\ref{fig:1}. Shifting it to the conical part of the spectrometer leads to larger axial and radial inhomogeneities of the electrical retarding potential and the magnetic field.
The variation of the electrical potential (magnetic field) across the analysing plane could reach $q\Delta U \approx {\cal O}(1)$\,eV ($\Delta B_\mathrm{ana}\approx {\cal O}$(0.1)\,mT) compared to $q\Delta U \approx 0.14$\,eV ($\Delta B_\mathrm{ana}\approx 0.004$\,mT) in the nominal setting, see Table~\ref{table:0}. This fact has further implications for the neutrino mass measurement (see section \ref{sec:new}).

It is important to notice, that besides reducing low energetic background from ionised Rydberg atoms, the SAP mode is expected to reduce the number of ionisations by the primary electrons from the short-lived $^{219}$Rn and $^{220}$Rn decays in the volume of the spectrometer. Due to the higher inhomogeneity of the electromagnetic field the trapping conditions for the highly energetic primary electrons become less favorable~\cite{ref:Gar58,ref:Roth64,ref:KT73,ref:Bieringer}. Besides, the magnetic mirror effect prevents part of the primary electrons to reach the downstream flux tube volume. These two effects additionally reduce of the number of secondary electrons reaching the detector from the smaller downstream flux tube volume. Therefore, the total amount of radon-induced background events will be reduced. Another benefit of the reduction of radon-induced background by the SAP setting is the fact that the number of time-correlated background events per one primary electron and, therefore, the non-Poisson overdispersion of the background count rate becomes significantly smaller.

\subsection{Field optimisations}
\label{sec:4.1}

The optimal SAP configuration is achieved by reducing the flux tube volume while keeping the inhomogeneity of the fields reasonably small and preserving the energy width of the transmission function. Special care has to be taken to avoid ``too early retardation'' and fulfill the requirement that the minimum of the kinetic energy component $E_\parallel$ parallel to the magnetic field is always the same regardless of the start pitch angle.
This requirement limits the position of the SAP. Another restriction is that the dimension of the spectrometer at larger $z$-position gets too small to guide the required flux tube if the magnetic field there is too small. 
The optimisation procedure contains two steps that are iterated afterwards: first the currents applied to the air coils are optimised with respect to the radial homogeneity at the minimum position, then the electric potential is modified to align the minima of the longitudinal kinetic energy. The spacial distance between the minima and the variation of electric potential across the analysing plane are minimized.

With the help of these extensive simulations of the fields an optimal position of the SAP was found at around 6\,m distance from the middle of the vessel towards the detector \cite{ref:Bieringer} (see Fig.~\ref{fig:1} (green)). The simulations are performed using an efficient field calculation software, benchmarked with the standard Kassiopeia package \cite{ref:Kassiopeia}. Table~\ref{table:0} summarises the main characteristics of the new SAP configuration compared to the symmetric 0.63\,mT setting.
In SAP configuration the downstream flux tube volume in Fig. \ref{fig:1} is reduced from $V_\mathrm{0.63\,mT} = 160$\,m$^3$ to $V_\mathrm{SAP}=56$\,m$^3$ so, that the Rydberg component of the background rate is expected to be reduced by a factor of $V_\mathrm{0.63\,mT}/V_\mathrm{SAP} \approx 2.8$, assuming a spatially homogeneous low-energetic background electron event distribution inside the main spectrometer.
\begin{table}[]
 \centering
\caption{ Comparison of the main parameters of the SAP and nominal 0.63\,mT configurations from simulation: $z$-position of the analysing plane relative to the middle of the spectrometer, volume of the downstream flux tube $V_\mathrm{down}$, magnetic field in the analysing plane $B_\mathrm{ana}$, variation of the electrical potential across the analysing plane $q\Delta U$ as a measure of the inhomogeneity, and the transmission width $\Delta E$.}  \begin{tabular}{c|c|c|c|c|c} 
   Setting  & $z$ (m) & $V_\mathrm{down}$ (m$^3$) & $B_\mathrm{ana}$ (mT) & $q\Delta U$ (eV) & $\Delta E$ (eV)  \\ 
   \hline
   SAP & $5.0-6.5$ & 56 & $0.45-0.62$ & 3 & $2.0-2.7$ \\ 
   \hline
   ``0.63\,mT'' & 0 & 160 & 0.63 & 0.14 & 2.8 \\ 
  \end{tabular}
  \label{table:0}
\end{table}

\begin{figure*}
 \centering
 \includegraphics[width=0.75\textwidth]{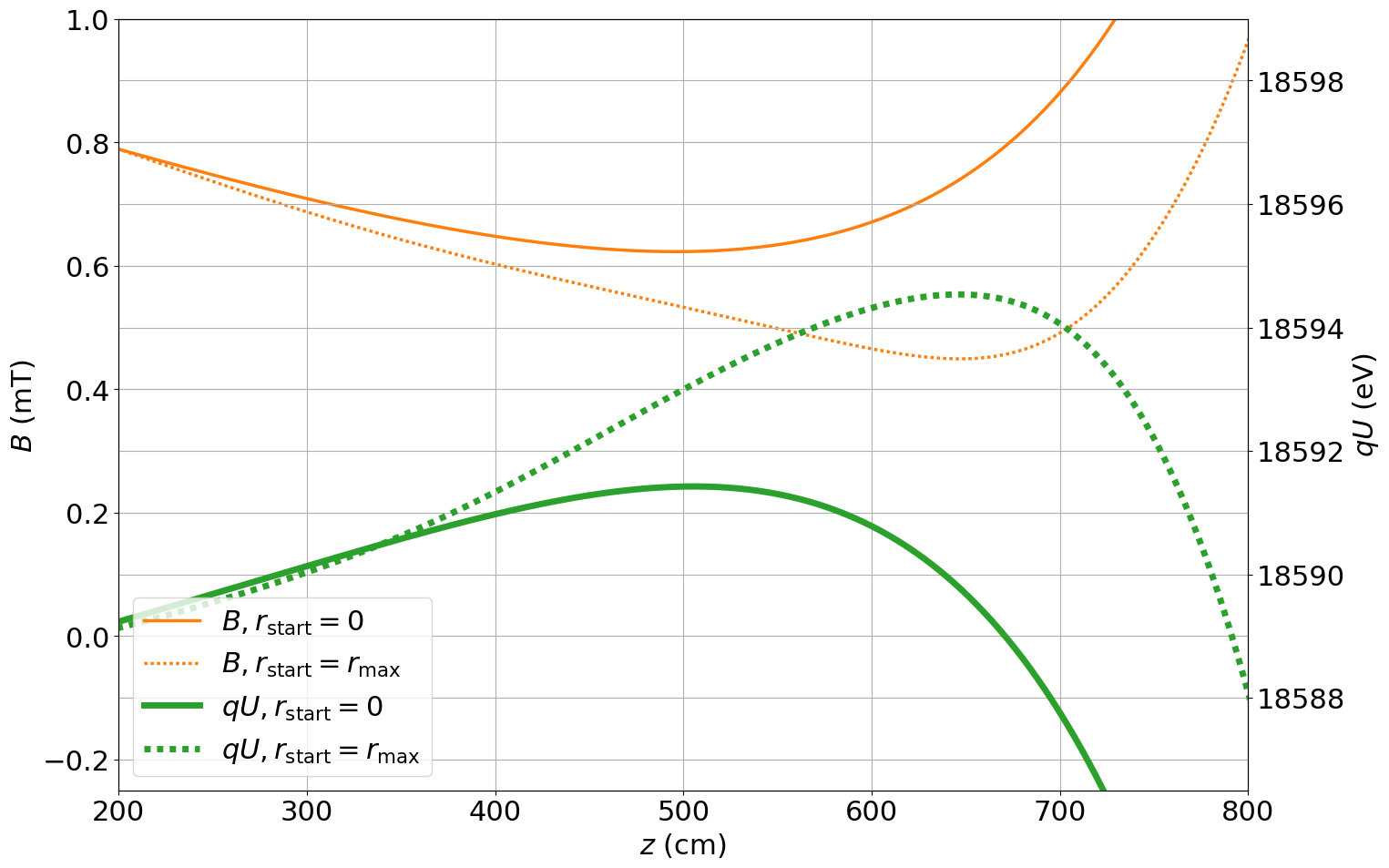}
 \caption{The retarding energy $qU$ and magnetic field near the analysing plane of the SAP configuration. The solid (dashed) lines show the $B$ (orange) and $qU$ (green) profiles for the inner (outer) part of the flux tube defined by $\Phi = 134\,\mathrm{T}\cdot \mathrm{cm}^2$. The minima of the magnetic field are aligned in $z$-position with the maxima of the retarding energy.}
 \label{fig:2}
\end{figure*}

Figure~\ref{fig:2} shows the matching of the minimum of the magnetic field and the maximum of the absolute retarding potential in the SAP configuration for electrons moving at different radii in the flux tube.  
%
%
\begin{figure*}
 \centering
 \includegraphics[width=0.75\textwidth]{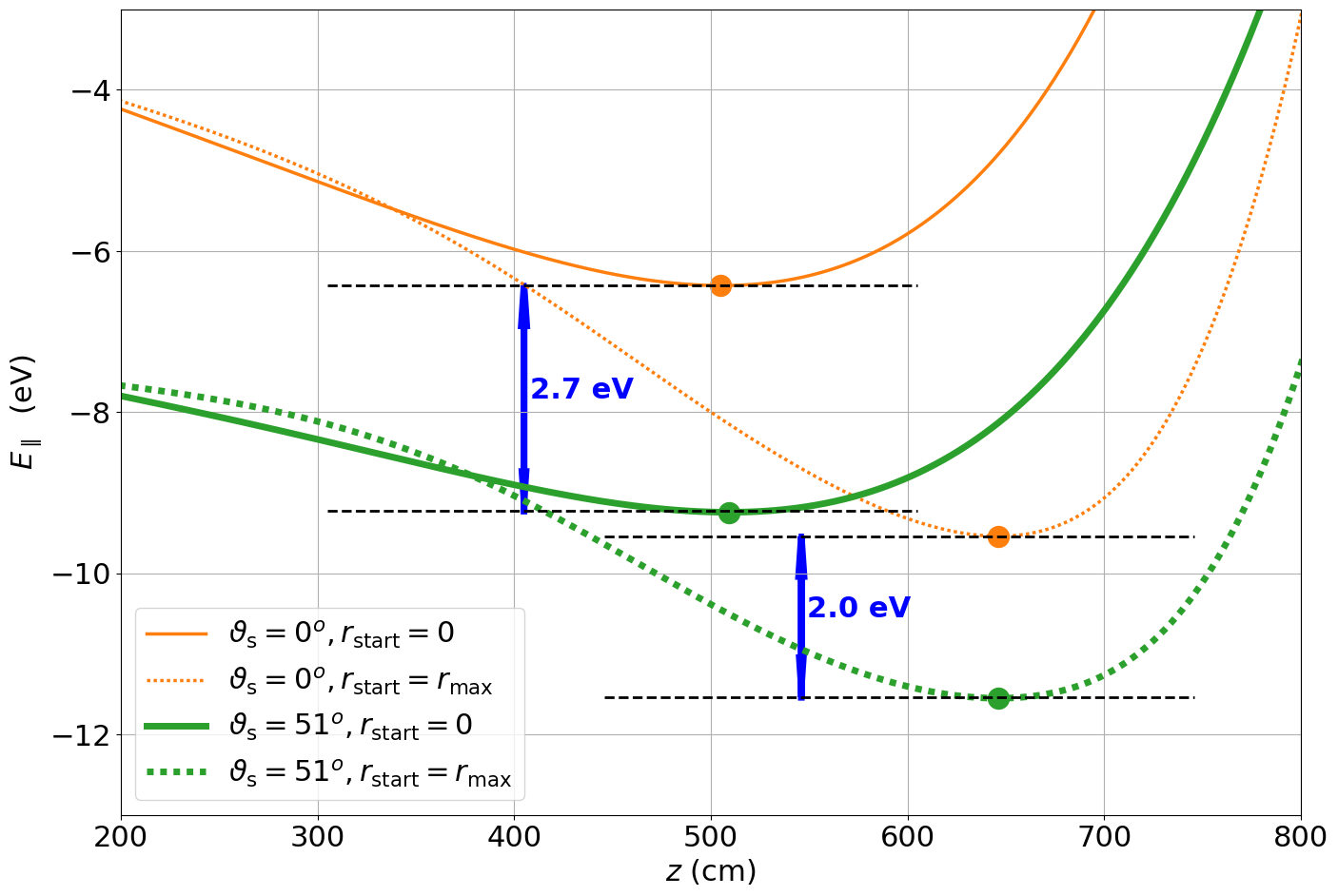}
 \caption{Longitudinal kinetic energy $E_\parallel$ of electrons (total energy of $E_\mathrm{tot}=18600$\,eV, $qU = -18615$\,eV) near the analysing plane in the SAP configuration for minimal ($\vartheta_\mathrm{s}=0^\circ$, thin orange lines) and maximal ($\vartheta_\mathrm{s}=\vartheta_\mathrm{max}=51^\circ$, thick green lines) pitch angle of the electrons.
 The minima of the longitudinal energy are coinciding within 5\,cm along the $z$-axis for electrons with different pitch angles but on the same magnetic field line. The difference of the minimal $E_\parallel$ values for the same radial position and minimal and maximal pitch angles determines the filter width of the spectrometer ($\Delta E = 2.7$ eV for the inner radii and $\Delta E = 2.0$ eV for the outer parts of the flux tube).}
 \label{fig:3}
\end{figure*}
Figure~\ref{fig:3} shows that the point of minimal longitudinal energy $E_\parallel$ of electrons starting on the same radius but with different pitch angles varies within an interval of about 5\,cm along the $z$-axis. This variation is larger than the design value of 5\,mm \cite{ref:wandkowski} and contributes an additional uncertainty of less than $5$\,meV to the filter width of the spectrometer.
The variation of the longitudinal energy $E_\parallel$ in the analysing plane for different starting angles of the electrons defines the transmission width of the spectrometer for an isotropic source.

\begin{figure*}
 \centering
 \includegraphics[width=0.6\textwidth]{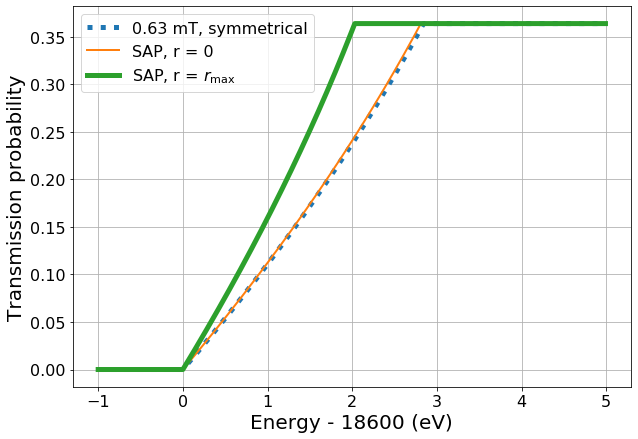}
 \caption{Transmission functions of the KATRIN main spectrometer for the SAP and symmetric configurations in the case of an isotropic source. The retarding energy is $qU$ = 18600\,eV, the magnetic fields are: 0.63\,mT for the symmetric setting (dotted, blue), 0.62\,mT for the inner part of the flux tube in SAP mode (dashed, orange), 0.45\,mT for the outermost radius ($r_\mathrm{max}\approx 2.8$~m) of the flux tube in SAP mode (solid, green). The transmission width of the latter two corresponds to the $\Delta E_\parallel$ in Fig.~\ref{fig:3}}.
 \label{fig:4}
\end{figure*}

Figure~\ref{fig:4} shows the transmission function of
the KATRIN spectrometer for the three cases: the symmetric configuration with 0.63\,mT magnetic field in the analysing plane averaged over all radii, the SAP configuration for electrons with $r=0$ and the SAP configuration for electrons in the outer radius of the flux tube. An isotropically emitting source and a maximum accepted start angle of $\vartheta_\mathrm{max} = \arcsin\left(\sqrt{\frac{B_\mathrm{source}}{B_\mathrm{max}}}\right)\approx 51^\circ$ at the source are assumed. The comparison of the three transmission functions shows that the SAP configuration provides even narrower transmission for the outer part of the flux tube while preserving the energy width of 2.7\,eV for the central part of the flux tube. The measured transmission for each of the detector pixels is given by a convolution of the transmission functions with the potential variation across the pixel.

The technical implementation of the SAP configuration is as follows. The rings of the inner electrode in the flat-cone part of the vessel on the detector side (ring numbers 12 - 14, Fig.~\ref{fig:0}) get the smallest positive offsets with respect to the other rings, see Table \ref{table:2}. 
\begin{table}[]
 \centering
\caption{Voltages applied to the inner electrode rings of the KATRIN main spectrometer in the SAP configuration. The common offset of the inner electrode system to the spectrometer vessel is -205\,V, the vessel voltage is -18400\,V. The voltage, applied to each IE-ring, is defined as $U_\mathrm{IE-ring} = U_\mathrm{vessel} + U_\mathrm{IE-common}+U_\mathrm{offset}$. The lowest voltage is placed at the middle ring of the flat cone part of the vessel (ring 13). Rings 12 and 14 are creating a shallow minimum of the electric potential in the $z$-direction. The geometry of the inner electrode system can be found in \cite{ref:TDR2}, it is shown schematically in Fig.~\ref{fig:0}.}
   \begin{tabular}{c|c|c|c|c|c|c|c|c} 
   IE-ring  & 2,3 & 4-6 & 7-11 & 12 & 13 & 14 & 15 & 16 \\ 
   \hline
   Offset, V & +125 & +25 & +10 & +2 & 0 & +2 & +45 & +105\\ 
   \hline
   Voltage, V & -18480 & -18580 & -18595 & -18603 & -18605 & -18603 & -18560 & -18500\\ 
  \end{tabular}
 \label{table:2}
\end{table}
The air coils in the source-side part of the spectrometer (coil numbers 1 - 10, Fig.~\ref{fig:0}) produce a relatively high magnetic field in this region to shift and shape the magnetic field minimum to the SAP position. One of the challenges of placing the analysing plane closer to the exit of the spectrometer is the stray field of the superconducting pinch magnet (4.2\,T) that defines the field near the beam axis. To counteract this field, several air coils (coil numbers 17 - 20) are producing a magnetic field in the opposite direction with their maximum allowed current of -120\,A, see Table~\ref{table:1}.
\begin{table}[]
 \centering
\caption{Currents applied to the air coils in the SAP configuration. The sign of the current defines the direction of the induced magnetic field w.r.t. the superconducting solenoids in the KATRIN beamline. Air coils with negative currents produce the magnetic field in the direction that is opposite to the field of the superconducting  solenoids. The geometry of the air coils can be found in \cite{ref:Erhard18,ref:TDR2}, it is shown schematically in Fig.~\ref{fig:0}.}
  \begin{tabular}{ c|c|c|c|c|c|c|c|c|c|c } 
  Air coil  & 1 & 2 & 3 & 4 & 5  & 6 & 7 & 8 & 9 & 10 \\ 
  Current, A & 120 & -4 & 116 & 70 & 70 & 70 & 70 & 110 & 110 & 110 \\ 
  \hline
  Air coil  & 11 & 12 & 13 & 14 & 15 & 16 & 17 & 18 & 19 & 20 \\ 
  Current, A & -60 & -28 & 58 & 5 & 120 & 120 & -120 & -119 & -120 & -120 \\ 
 \end{tabular}
 \label{table:1}
\end{table}


\subsection{Measurement results}
\label{sec:4.2}

The principle of background reduction by an SAP configuration was demonstrated already in 2015 during the commissioning of the KATRIN experiment \cite{ref:Dyba}. After other background sources were successfully mitigated (see e.g. references~\cite{ref:penning,ref:goerhardt}) and the Rydberg background was better understood \cite{ref:Rydberg} extensive investigations to determine an optimal SAP configuration were performed \cite{ref:Bieringer,ref:Schaller}. 

The tests described in this paper were performed in autumn 2019 after the first science run of KATRIN and after a bake-out of the spectrometer vessel, which had reduced the background level by about 30\%. The main aim of the measurements was to choose an optimal configuration of the electromagnetic field that provides a substantial (factor 2) reduction of the measured background rate while preserving the transmission properties (e.g. transmission width below 2.8~eV) for the neutrino mass measurements.

During these studies the background rate was measured in intervals of 30 minutes to several hours for a series of predefined sets of air coil currents and electric potentials of the inner electrode rings. The settings cover a wide range of volumes of the downstream flux tube, $1-656$\,m$^3$, and different variations of the electric potential and magnetic field across the analysing plane. The settings were chosen according to the simulations of the flux tube and transmission properties in the main spectrometer using dedicated software \cite{ref:Bieringer}.

\begin{figure*}
 \centering
 \includegraphics[width=0.75\textwidth]{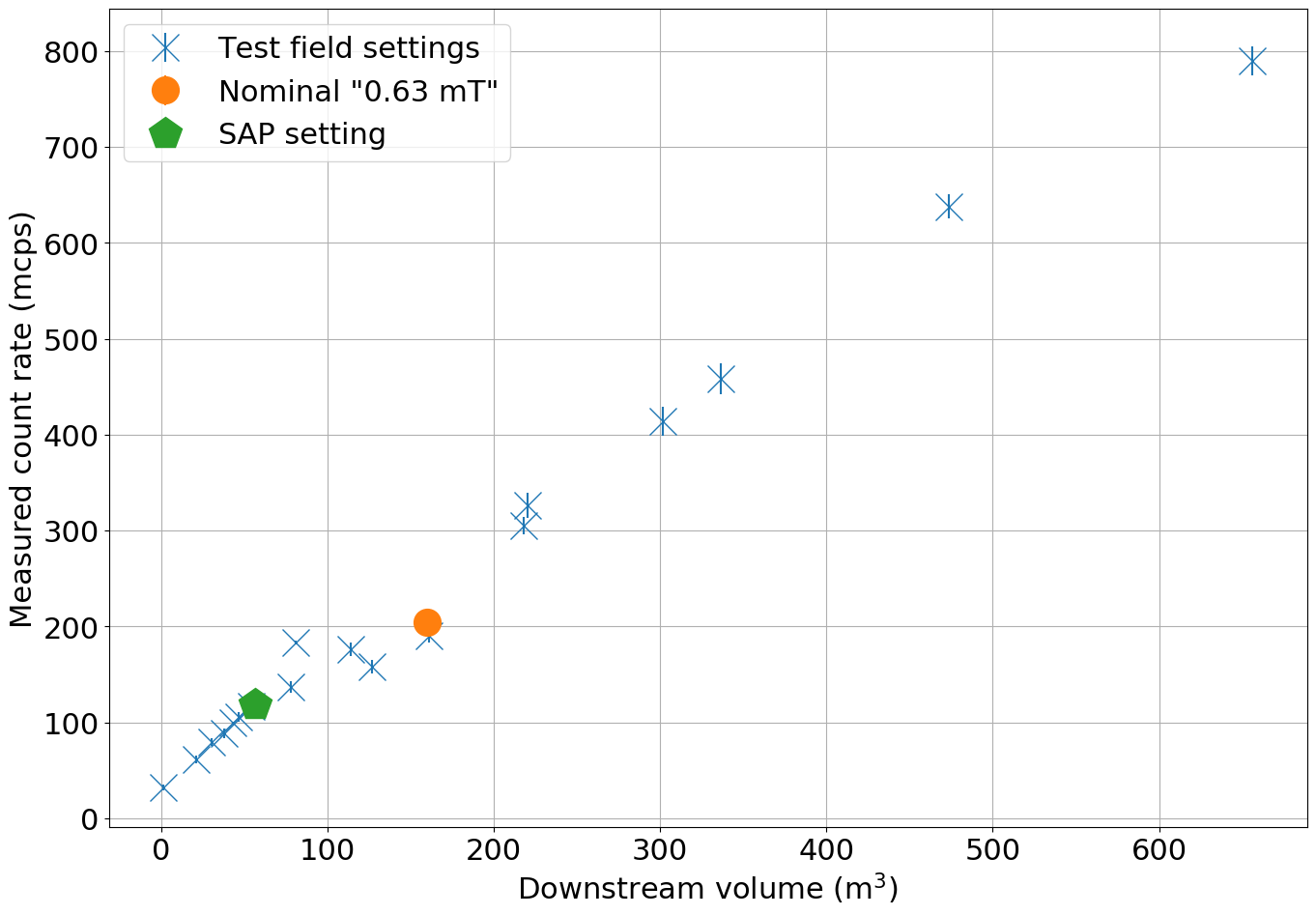}
 \caption{Measured background rate as function of the calculated downstream flux tube volume.
 The blue crosses show the measured rate for several field configurations with various flux tube volumes.
 The orange (large circle) and green (pentagon) data points show the background rates for the nominal 0.63\,mT and the SAP settings, respectively. The configuration of the green (pentagon) data point is optimal with respect to the background rate and the homogeneity of the electromagnetic field. Average detector related background rate of 35 mcps is not subtracted. Statistical uncertainties for some points are to small to be visible.} 
 \label{fig:5}
\end{figure*}

Figure~\ref{fig:5} shows the measured background rates as a function of the estimated
volume of the downstream flux tube. 
The measured rates exhibit a nearly linear dependence on the downstream flux tube volume. Deviations from this behaviour might be due to an inhomogeneous distribution of the starting points of the Rydberg electrons \cite{ref:Bieringer,ref:Hinz} or other background processes like trapped electrons from short-lived $^{219}$Rn and $^{220}$Rn decays.


Several configurations show a clear reduction of the spectrometer background rate (apart from the detector intrinsic background of $\sim 35$\,mcps). However, a trade-off is required between the remaining background rate, the filter width $\Delta E$ and the homogeneity of the electric and magnetic fields over the analysing plane to keep systematic uncertainties introduced by the SAP configuration at an acceptable level.


The optimal SAP configuration described in section \ref{sec:4.1} (Fig.~\ref{fig:5}, green pentagon) possesses a downstream flux tube volume of about 56\,m$^3$ (compared to a downstream flux tube volume of 160\,m$^3$ in the nominal configuration). The measured total background count rate in this SAP configuration was determined to be ($120\pm1$)\,mcps (including $\sim 35$\,mcps of intrinsic detector background rate) compared to a total background rate of ($204\pm1$)\,mcps in the nominal configuration at the time of the measurement shortly after the bake-out of the spectrometer vessel. In conclusion a factor of 2 reduction of the volume-related background rate is achieved in the SAP operating mode.


\section{Implications for the neutrino mass measurement}
\label{sec:new}

As described above, the SAP configuration enables a background reduction of a factor of 2, however, it leads to a much larger magnetic field and electric potential variations across the analysis plane than in the symmetric case. These variations have two main consequences for the neutrino mass analysis.

First, in order to take into account the inhomogeneities of the electric potential and the magnetic field in the analysing plane of the SAP settings, the analysis exploits the radial and azimuthal pixelization of the detector \cite{ref:FPD}. The detector is segmented in 148 pixels, arranged in concentric rings. Each pixel observes a statistically independent tritium beta decay spectrum. 
In an ideally aligned system, all detector pixels on one ring would be combined, thus minimising the inhomogeneities. In the real KATRIN experiment with small misalignments, groups of pixels with almost the same electrical potential and almost identical magnetic field defining ``patches'' can be analysed together. The magnetic field and electric potential variation over a patch amounts to less than $\delta B \leq 0.014$\,mT and $\delta qU \leq 0.25$\,eV, respectively. In the neutrino mass analysis these detector patches are fitted simultaneously, but with individual model predictions taking into account the patch-dependent fields, and thus enhancing the needed computation power. As a proof of concept, it was successfully implemented for the second neutrino mass measurement data analysis \cite{ref:KNM2} and also tested in a new model calculation technique using a neural network \cite{ref:Netrium}.

The second consequence of the SAP configuration for the neutrino mass measurement is the need for a precise experimental determination of the magnetic fields and electric potential over the analysing plane. In contrast, in earlier measurements using the symmetric analysing plane the potentials and fields in the analysing plane were based on simulations and the electric potential variation over a single patch could be neglected.

A precise measurement of the fields is possible, for instance, with a gaseous krypton source \cite{ref:KrCalibration}. By scanning narrow lines of $^\mathrm{83m}$Kr (for instance, K-32, L$_3$-32 and N$_{2,3}$-32 lines) with the MAC-E filter the electric potential, the magnetic field and the effective broadening of the transmission due to $\delta B$ and $\delta qU$ over a patch can be measured directly. 
Such a measurement was performed and meets the KATRIN design requirement regarding the systematic contribution of the SAP fields to $m_\nu^2$ on the level of $0.0075$~eV$^2$. The details of this approach will be described in a separate publication \cite{ref:KrCalibration,ref:Block}.


\section{Conclusion and outlook}
\label{sec:5}
To reach the target sensitivity of the KATRIN experiment to the neutrino mass of 0.2\,eV$/c^2$ (at 90\%\,CL), a reduction of the background rate to values below 0.1\,cps is required.

The novel shifted analysing plane (SAP) configuration described in this paper reduces the background rate from the KATRIN main spectrometer by a factor of 2, allowing tritium beta-decay measurements at a background rate of 0.12\,cps.
The configuration provides a narrow filter width of the KATRIN main spectrometer in the range of 2.0\,eV to 2.7\,eV at an electron energy of 18.6\,keV, surpassing the width of 2.8\,eV in the symmetric configuration (0.63\,mT) used in the neutrino mass measurements so far.

The minimal magnetic field is varying in the SAP configuration from 0.45\,mT to 0.62\,mT over the analysing plane and the corresponding electric potential variation is 3\,eV. 
To measure the electromagnetic field in-situ a precise calibration procedure was developed and performed using monoenergetic conversion electrons from a $^\mathrm{83m}$Kr source~\cite{ref:FirstLight,ref:Krypton} and, in addition, monoenergetic electrons from an angular-selective photo-electron source~\cite{ref:egun,ref:TDR2}. Along with the determination of the electromagnetic field, test scans of the tritium spectrum were performed in the SAP configuration to assure that the neutrino-mass data-taking is not affected by unexpected side-effects of this measurement mode.  
The corresponding neutrino mass analysis makes full use of the detector pixelization. The additional systematic uncertainties that occur are marginal \cite{ref:KrCalibration,ref:Block}.
After the careful calibration the shifted analysing plane was implemented as the new default beta-spectrum scanning mode in KATRIN since 2020.




Several other options of detecting or removing the remaining part of the KATRIN background, Rydberg background electrons, are considered, including time-of-flight spectroscopy \cite{ref:tof}, the idea of time-focusing time-of-flight measurements \cite{ref:tftof} or making use of the specific angular distribution of the Rydberg electrons to filter out background events.
Should one of these methods be successful, a symmetric configuration of the fields might be considered again because of the improved filter width (e.g. with minimal magnetic field of 0.21\,mT reaching the design value $\Delta E = 0.93$\,eV) and reduced systematics related to the homogeneity of the electromagnetic field. 

\begin{acknowledgements}

We acknowledge the support of Helmholtz Association (HGF), Ministry for Education and Research BMBF (05A17PM3, 05A17PX3, 05A17VK2,05A17PDA, 05A17WO3 and 05A20PMA), Helmholtz Alliance for Astroparticle Physics (HAP), the doctoral school KSETA at KIT, and Helmholtz Young Investigator Group (VH-NG-1055), Max Planck Research Group (Max-Planck@TUM), and Deutsche Forschungsgemeinschaft DFG Research Training Groups Grants No., GRK 1694 and GRK 2149, Graduate School Grant No. GSC 1085-KSETA, and SFB-1258 in Germany; Ministry of Science and Higher Education of the Russian Federation under contract 075-15-2020-778. This project has received funding from the European Research Council (ERC) under the European Union Horizon 2020 research and innovation programme (grant agreement No. 852845).

\end{acknowledgements}





\end{document}